%% file: barr.tex
\documentclass[]{aa}

\usepackage{txfonts,graphicx}

\newcommand{\arcsd}[2]{$#1^{\prime\prime}\!\!.#2$}

\newcommand{\etal}{{et al.} }
\newcommand{\eg}{{e.g. }}
\newcommand{\ie}{{i.e. }}

\newcommand{\afe}{$[\alpha/\mathrm{Fe}]$}
\newcommand{\fe}{$\langle \mathrm{Fe} \rangle$}
\newcommand{\feh}{$[\mathrm{Fe}/\mathrm{H}]$}

\newcommand{\cosmology}{$H_0 = 70$ km s$^{-1}$ Mpc$^{-1}$,
$\Omega_{\mathrm{m}} = 0.3$, $\Omega_{\Lambda} = 0.7$}

\begin{document}

\title{The formation of S0 galaxies: evidence from globular clusters}

\author{J.~M.~Barr\inst{1} \and A.~G.~Bedregal\inst{1} \and A.~Arag{\'o}n-Salamanca\inst{1} \and M.~R.~Merrifield\inst{1} \and S.~P.~Bamford\inst{1,2}}

\titlerunning{The formation of S0 galaxies}
\authorrunning{J.~M.~Barr et al.}

\institute{The School of Physics \& Astronomy, University of Nottingham, University Park, Nottingham, NG7 2RD, UK \and Institute of Cosmology and Gravitation, Mercantile House, Hampshire Terrace, University of Portsmouth, Portsmouth, PO1 2EG, UK}

\date{Received / Accepted }

\abstract
{} 
{We devise a simple experiment to test the theory that lenticular (S0)
galaxies form from spirals whose star formation has been shut down. An
individual galaxy's fading is measured using the globular cluster
specific frequency ($S_N$), defined as the number of globular clusters
normalised by the galaxy luminosity. This is compared with a
spectroscopically-derived age estimate.}
{We make NTT/EMMI long-slit spectroscopic observations of 11 S0
galaxies at $z < 0.006$. We measure the absorption-line indices,
H$\delta$, H$\gamma$, Mg$b$, Fe5270 and Fe5335 within the central
$r_e/8$. By inverting single-stellar population models,
luminosity-weighted mean ages, metallicities and $\alpha$-element
abundance ratios are derived. We estimate the amount of fading a
galaxy has undergone by comparing each galaxy's $S_N$ with its
deviation from the mean spiral $S_N$.}
{Galaxies with higher $S_N$ have older stellar populations. Moreover,
we find that the zero-point and amount of fading is consistent with a
scenario where lenticulars are formed by the quenching of star
formation in spiral galaxies. Our data also rule out any formation
method for S0s which creates a large number of new globular
clusters. We confirm that previous results showing a relationship
between $S_N$ and color are driven by the $S_N -$ Age relation. Five
galaxies show detectable H$\beta$, [O{\sc~iii}], H$\alpha$ or
[N{\sc~ii}] emission lines. However, only in the two youngest galaxies
is this emission unambiguously from star formation.}
{Our results are consistent with the theory that S0 galaxies are
formed when gas in normal spirals is removed, possibly as a result of
a change in environment. The on-going star formation in the youngest
galaxies hints that the timescale of quenching is $\lesssim 1$ Gyr. We
speculate, therefore, that the truncation of star formation is a
rather gentle process unlikely to involve a rapid burst of star
formation.}


\maketitle

\section{Introduction}

Lenticular (S0) galaxies live at the intersection of spirals and
ellipticals on Hubble's Tuning Fork. As a class, they provide a useful
exemplar of what could be an intermediate stage of a galaxy's
evolution. They also give us insight into galaxy formation and its
relationship with environment. Hubble Space Telescope observations of
distant galaxy clusters show that the proportion of S0s declines with
redshift, while the abundance of spirals increases
\cite{dressler97}. The idea that star formation in spiral galaxies is
cut off when they enter a denser environment therefore seems a
plausible one. The mechanism by which this cessation is achieved is a
topic of active debate and many scenarios have been proposed. These
include close encounters or mergers, which increase the luminosity of
the bulge component by heating the central parts of the disk or
triggering a central star-formation episode (\eg Mihos \& Hernquist
1994; Bekki 1998)\nocite{mihos94,bekki98}. Galaxy harassment, where a
galaxy undergoes many close but fleeting high-speed interactions with
other galaxies (\eg Moore \etal 1996, 1998)\nocite{moore96,moore98}
is predicted to have a similar effect. Other scenarios involve the
interaction of a spiral galaxy with the intra-cluster gas either by by
ram-pressure stripping (\eg Gunn \& Gott 1972\nocite{gunn72}; Quilis
\etal 2000\nocite{quilis00}; Vollmer \etal 2001\nocite{vollmer01}; Sun
\etal 2006\nocite{sun06}), or over a longer period, for example by
removal of gas from the galaxy halo (\eg Larson \etal
1980\nocite{larson80}), or by heating of gas within the galaxy by the
ICM -- so-called thermal evaporation \cite{cowie77}. See Bosselli \&
Gavazzi (2006) for a thorough discussion of the various mechanisms.

These transformation scenarios can be separated in a number of ways,
perhaps contrasting the effects of other galaxies against intracluster
gas, or looking at gravitational versus hydrodynamical
drivers. However, from an observational point of view the most
accessible information is how rapid or violent a particular
transformation is. Even if S0 galaxies are formed via a unique
mechanism, {\it post hoc} observations will not easily distinguish
between those transformations which yield a similar final state. As
well as looking for consistency with a fading scenario, this study
will test whether lenticular formation is more likely to be due to a
violent or a passive episode.

Direct observational studies of S0 formation, as opposed to
simulations, are rather thin on the ground. Dressler et
al. (2004)\nocite{dressler04} show evidence from composite spectra in
rich clusters that S0 galaxies probably experienced a recent burst of
star formation. It has also been suggested that the mechanism of
truncation was more violent at higher redshift (see Boselli \etal
2006)\nocite{boselli06}.

Circumstantial evidence in support of quiet lenticular formation comes
from observations of the S0 Tully-Fisher relation
\cite{bedregal06}. In the $B$-band this is, on average, $\sim 1.3$ mag
fainter than the spiral relation of Sakai \etal
(2000)\nocite{sakai00}. The scatter is also much larger. This can be
interpreted as a fading of a stellar population of a given rotational
velocity, where the fading begins over a range of epoch, corresponding
to the cessation of a galaxy's star formation. This does, however,
presume that the rotational velocity of galaxy is not greatly altered
in a transformation from spiral to S0. And indeed that you can
accurately disentangle the rotational velocity from velocity
dispersion in an S0 using the combination as a proxy for mass. This
result also relies on the fact that the progenitors of the current
generation of S0s can be compared directly with local spirals. Were
there to be strong evolution in the masses of spiral galaxies since $z
\sim 0.5$ this comparison would not be valid. There is little
consensus at present on whether this is the case; see Flores \etal
(2006)\nocite{flores06} and Weiner \etal (2006)\nocite{weiner06} for
contrasting viewpoints.

A simple observational test between formation mechanisms is provided
by globular clusters. It is widely presumed that globular clusters are
created and disrupted during the kind of violent galaxy interaction
like a merger (\eg Ashman \& Zepf 1998\nocite{ashman98}). In contrast,
if we assume that a quieter scenario (\eg ram-pressure stripping,
thermal evaporation) will roughly preserve the number, and luminosity,
of a galaxy's globular clusters, we can use the specific frequency of
globular clusters ($S_N$) as a diagnostic tool. Specific frequency is
defined as the number of globular clusters per unit luminosity in a
galaxy. If star formation shuts down in a spiral galaxy, $S_N$ will
increase as the galaxy fades. This quantity can therefore be used to
trace the time since the last star formation episode. If the passive
mechanism holds, then we expect to see a correlation between stellar
age and $S_N$ in S0s. No such correlation is expected in mergers and
close gravitational encounters as these are not expected to preserve
$S_N$.

Arag{\'o}n-Salamanca \etal (2006)\nocite{aragon-salamanca06} examined
color as a function of $S_N$ for a sample of 12 S0 galaxies. They used
color as a proxy for age and found that $U,B,V,R,I$ colors are related
to $S_N$ in the manner expected of a fading stellar population. That
is to say the redder galaxies have higher $S_N$, \ie have undergone
more fading. Moreover, the distribution of colors and $S_N$ is
entirely consistent with models predicting the color and luminosity
evolution of single-stellar populations (\eg Worthey 1994; Bruzual \&
Charlot 2003)\nocite{bruzual03,worthey94}. Furthermore, comparing
values of $S_N$ with spirals indicated that S0s had faded by a factor
of $\sim 3$, the same amount found by Bedregal \etal
(2006)\nocite{bedregal06} for the S0 Tully-Fisher relation.

The problem that observations of color and luminosity encounter is the
age-metallicity degeneracy. It has not been possible to say for
certain that the physical effect driving the color-$S_N$ relation is
age. Clearly, what is needed is a test that can separate age and
metallicity, and ideally mass and luminosity as well. In this paper we
use spectroscopic observations of absorption-line strengths to derive
physical attributes of a sample of S0s and compare these with
$S_N$. Section~\ref{sec:dat} describes the observations and data
reduction. Section~\ref{sec:res} gives results. We work towards
putting our galaxies on the Age $-S_N$ plot in
Section~\ref{sec:dis}. Conclusions are presented in
Section~\ref{sec:con}. Throughout we use \cosmology.

\section{Observations and data reduction}
\label{sec:dat}

\subsection{Sample selection}

The 11 lenticular galaxies in our sample are a subset of those with
globular cluster observations in Kundu \& Whitmore
(2001b)\nocite{kundu01b}. The galaxies are chosen to cover a range in
luminosity and globular cluster specific frequency. All apart from two
(NGC 3056 and NGC 3115B) reside in group or poor cluster
environments. See Kundu \& Whitmore (2001b)\nocite{kundu01b} for more
details.

\subsection{Globular cluster specific frequency}
\label{sec:glo}

The specific frequency of globular clusters is defined as,$$ S_N = N_t
10^{0.4(M_V +15)} $$ where $N_t$ is the total number of clusters and
$M_V$ is the total $V$-band magnitude of the galaxy\footnote{Note that
we fix the error in Ashman \& Zepf (1998)\nocite{ashman98} where $S_N$
is defined with a negative power. See Harris \& van den Bergh
(1981)\nocite{harris81}.}.

HST-derived values of local $S_N$ for S0s come from Kundu \& Whitmore
(2001b)\nocite{kundu01b}. Local in this case refers to the fact that
the WFPC2 field-of-view only covers the central part of the galaxy and
so the number of clusters and normalising luminosity are not those of
the galaxy as a whole. The ratio of local to global $S_N$ depends to
some extent on luminosity, though the uncertainties are high. See
Kundu \& Whitmore (2001a,b) for more details. We use the
directly-derived, local values of $S_N$. There is also a closer
correspondence of scales between these values and our spectral
aperture. We note that the average local values of $S_N$ for spirals
is 0.4 \cite{goudfrooij03}, lenticulars 1.0 \cite{kundu01b}, and
ellipticals 2.4 \cite{kundu01a}.

\subsection{EMMI Data}

Spectroscopic observations of our targets was obtained on the NTT at
La Silla during 2005 December 1--3. Observations were made using EMMI
\cite{dekker86} in its long-slit, low-resolution grism spectroscopy mode
(RILD). Seeing varied from \arcsd{0}{45} to \arcsd{1}{4}
FWHM. Instrumental parameters are given in
Table~\ref{tab:ins}. Properties of the targets are detailed in
Table~\ref{tab:obs}.

\begin{table}

  \caption{Instrument Parameters\label{tab:ins}}
  \centering

  \begin{tabular}{lr}
  
    \hline\hline
    \\		
    Telescope & NTT \\
    Instrument & EMMI \\
    Wavelength range & $3500-7200$\AA \\
    Grism & LR5 \\
    Slit width & \arcsd{1}{02} \\
    Spatial resolution & \arcsd{0}{33} pix$^{-1}$ \\
    Spectral resolution$^{*}$. & 119 km s$^{-1}$ \\
    \\
    \hline\\

  \end{tabular}

  $^{*}$ Median instrumental resolution at 5500\AA.

\end{table}

\begin{table}
\caption{Targets and exposure times\label{tab:obs}}

\begin{tabular}{lrrr@{$\pm$}rrr}

  \hline\hline\\

  \multicolumn{1}{c}{Galaxy} & \multicolumn{1}{c}{T} &
  \multicolumn{1}{c}{$M_V$} & \multicolumn{2}{c}{$S_N$} &
  \multicolumn{1}{c}{$r_e$} & \multicolumn{1}{c}{$z$} \\

  & \multicolumn{1}{c}{(s)} & & \multicolumn{2}{c}{} &
  \multicolumn{1}{c}{($^{\prime\prime}$)} & \\
 
  \\

   \hline\\

   \input{tab1.asc}

   \\
   \hline\\

   \end{tabular}

  $M_V$ and $S_N$ from Kundu \& Whitmore
  (2001b)\nocite{kundu01b}. Half light radius from RC3 catalogue
  \cite{corwin04} except IC1919 where $r_e$ estimated from light
  profile in slit.

\end{table}

Data is reduced in standard fashion. The spectra are bias-subtracted
and flat-fielded using lamp flats. Wavelength calibration is performed
using arc spectra from the same night. We check the accuracy of the
solution using transformed arc spectra. The offset between measured vs
actual wavelength for a line increases linearly with distance from the
blaze wavelength at 5300\AA. This typically reaches a maximum of
0.1\AA \ at 4000\AA \ and 7000\AA.

The sky is removed by fitting a 3rd order Legendre polynomial to a
100$^{\prime\prime}$ aperture either side of the galaxy. The spectra
are extracted from an aperture of width equal to 1/8th of the
half-light radius. Flux calibrations are made using observations of
spectrophotometric standard White Dwarfs, G158-100, Feige 110, GD50
and GD108 undertaken during the same run.

To derive accurate kinematic information, observations of three
(K2III, G0V, F5V) template stars are made. These are reduced in
exactly the same way as the science data.

\subsection{Kinematics and Absorption-line Indices}

Velocity dispersions ($\sigma$) are derived using the method of
Gebhardt et al. (2000)\nocite{gebhardt00} and checked for consistency
with the code made available by Michele
Cappellari\footnote{http://www.strw.leidenuniv.nl/$\sim$mcappell/idl/}
(see Cappellari \& Emsellem 2004)\nocite{cappellari04}. Both methods
gave consistent results. Seven galaxies have $\sigma \lesssim $ the
instrumental resolution. This means they could be subject to
systematic uncertainties. However, velocity dispersions are only used
to provide a correction for the line strength measurements to the rest
frame. Errors that may be introduced in this way are small; for
example, a factor of two overestimate in the value of $\sigma$ causes
errors of $\sim 3\%$ in the H$\beta$ index and $\sim 5\%$ in
Mg$b$. These are of the same order as the random errors.

Absorption line indices are derived from the spectra. We use Mg$b$ and
\fe \ (the mean of Fe5270, Fe5335) from the Lick/IDS system as
defined in Worthey et al. (1994)\nocite{worthey94}. In addition we
calculate the H$\beta_\mathrm{G}$ index as in J{\o}rgensen
(1997)\nocite{jorgensen97} and the H$\gamma$ and H$\delta$ indices of
Worthey \& Ottaviani (1997)\nocite{worthey97}. Line indices are
corrected for the effect of velocity dispersion following the
technique described in Davies \etal (1993)\nocite{davies93}. We use the
models of Thomas et al (2003, 2004)\nocite{thomas03,thomas04} to
assign relative ages, metallicities and $\alpha$-element abundance
ratios (\afe) from the measured index values. Derived properties for
the sample are listed in Table~\ref{tab:spe}.

\begin{table*}
  \caption{Derived spectroscopic parameters\label{tab:spe}}

  \begin{tabular}{lr@{$\pm$}lr@{$\pm$}lr@{$\pm$}lr@{$\pm$}lr@{$\pm$}lr@{$\pm$}lr@{$\pm$}lr@{$\pm$}l}

  \hline\hline\\

  \multicolumn{1}{c}{Galaxy} & \multicolumn{2}{c}{$\sigma$} &
  \multicolumn{2}{c}{H$\gamma$} & \multicolumn{2}{c}{H$\delta$} &
  \multicolumn{2}{c}{Mg$b$} & \multicolumn{2}{c}{\fe} &
  \multicolumn{2}{c}{log Age} & \multicolumn{2}{c}{\feh} &
  \multicolumn{2}{c}{\afe} \\

  & \multicolumn{2}{c}{ (km s$^{-1}$) } & \multicolumn{2}{c}{ (\AA) }
  & \multicolumn{2}{c}{ (\AA) } & \multicolumn{2}{c}{ (\AA) } &
  \multicolumn{2}{c}{ (\AA) }  & \multicolumn{2}{c}{ (log Gyr) } \\
 
  \\

   \hline\\

    \input{tab2.asc}

    \\
    \hline\\

  \end{tabular}

  Stars denote those galaxies with line emission in H$\beta$,
  [O{\sc~iii}] [N{\sc~ii}], H$\alpha$ or [S{\sc~ii}].

\end{table*}

\section{Results}
\label{sec:res}

\subsection{Balmer lines, H$\beta$ vs H$\gamma$ and H$\delta$}

Traditionally, in trying to quantify age and metallicity using
spectroscopy, the H$\beta$ index is used as the primary age indicator
together with the Mg$b$ and \fe \ indices. An alternative,
particularly useful at higher redshift is to use a combination of the
higher order Balmer lines, H$\gamma$ and H$\delta$. However, it has
been noted by various investigators (\eg Terlevich \etal 1999;
Poggianti \etal 2001; Kuntschner \etal
2002)\nocite{terlevich99,poggianti01,kuntschner02} that H$\beta$ tends
to give systematically older ages than H$\gamma +$ H$\delta$.

There are two complementary but quite different effects responsible
for this discrepancy. Firstly, H$\gamma$ and H$\delta$ are more
sensitive to the effects of enhanced \afe. Derivations of ages for
early type galaxies using these indices which do not account for \afe
\ lead to younger ages than through H$\beta$ (see Thomas \etal
2004\nocite{thomas04}). Secondly, unresolved emission can ``fill in''
the stellar absorption line, causing it to appear weaker than it
otherwise would. In galaxies with relatively little on-going star
formation, this effects the lower-excitation states more than the
higher. This means that young galaxies could appear artificially old
when using H$\beta$ as the age diagnostic rather than the higher-order
Balmer lines.

We correct for the first phenomenon using models which include variable
\afe. To gauge the effect of unresolved emission, we note that, of our
sample, 2 galaxies have ostensible emission in H$\beta$. Three more
have [O{\sc~iii}], H$\alpha$ or [N{\sc~ii}]. Moreover, if we plot the
Balmer lines against one another we find a strong correlation between
H$\gamma$ and H$\delta$ but only a weak one between H$\beta$ and
either H$\gamma$ or H$\delta$. For these reasons, we conclude that
H$\beta$ is affected by unresolved emission in several cases. We
therefore use the H$\gamma +$ H$\delta$ index as the primary age
indicator in our analysis.

\subsection{Separating age, metallicity, and \afe}

\begin{figure}

  \includegraphics[width=80mm]{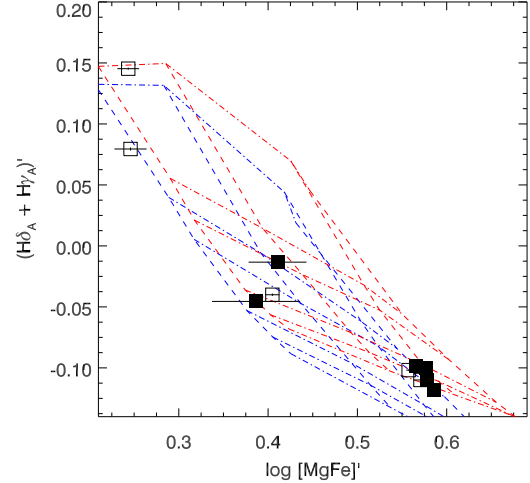}

  \caption{The higher order Balmer lines versus the Thomas \etal
  [MgFe]$^{\prime}$ index. Open points are those galaxies with either
  H$\alpha$, [O{\sc~iii}] or H$\beta$ emission. Grids show the
  Age-metallicity relations for \afe=0.0 (blue) and \afe=0.3
  (red). Dashed lines are (left to right) \feh=-0.3, 0.0, 0.3,
  0.5. Dot-dashed lines are ages of (top to bottom) 1, 2, 4, 8, 11, 12
  Gyr.}

\label{fig:hdg}

\end{figure}

\begin{figure}

  \includegraphics[width=80mm]{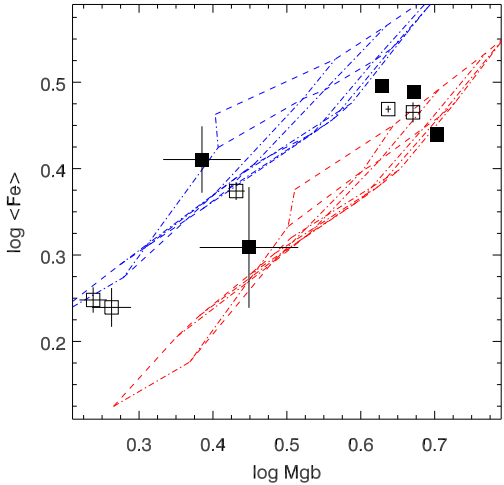}

  \caption{The Mg$b$ index versus \fe. Points and grids are equivalent to
  those in Figure~\ref{fig:hdg}.\label{fig:mgb}}

\end{figure}

In Figure~\ref{fig:hdg} we plot $(\mathrm{H}\delta_A +
\mathrm{H}\gamma_A)^{\prime}$ against the combination of indices
[MgFe]$^{\prime}$ of Thomas \etal
(2003)\nocite{thomas03}\footnote{$(\mathrm{H}\delta_A +
\mathrm{H}\gamma_A)^{\prime} = -2.5 \log [ 1 - (\mathrm{H}\delta_A +
\mathrm{H}\gamma_A)/82.5]$; $[\mathrm{MgFe}]^{\prime} = (\mathrm{Mg}b
\cdot [ 0.72 \cdot \mathrm{Fe}5270 + 0.28 \cdot \mathrm{Fe5335}
])^{0.5}$.}. This quantity, when plotted against H$\beta$, is designed
to be independent of \afe. However, when plotted against the
higher-order Balmer lines, the lack of correspondence in the
age-metallicity grids shows that some degeneracy remains. For this
reason we are not able to assign unique ages from this figure alone,
though our sample does clearly have a range of luminosity-weighted
mean ages and metallicities. It is therefore instructive to plot Mg$b$
vs \fe \ in Figure~\ref{fig:mgb}. This can be thought of as orthogonal
to Figure~\ref{fig:hdg}. We can see immediately that over half the
sample have super-solar \afe. Values of \afe \ for galaxy cluster
elliptical galaxies typically range over 0 -- 0.25 \cite{thomas05}.

\nocite{thomas03}

We now invert the models to assign meaningful physical quantities to
the galaxies. For each galaxy we fit $\mathrm{H}\delta_A +
\mathrm{H}\gamma_A$, Mg$b$, and \fe \ simultaneously and linearly
interpolate between points on the model grid. The values of age,
metallicity and \afe \ that minimise $\chi^2$ are adopted. Errors are
estimated using a Monte-Carlo method. Each combination of
$\mathrm{H}\delta_A + \mathrm{H}\gamma_A$, Mg$b$, and
\fe \ is perturbed with a Gaussian probability of
$\sigma_{\mathrm{index}}$ 1000 times assuming that measurements of the
indices are independent. Errors in ages, \feh \ and \afe \ are quoted
as the $68$th percentile of the distribution.

Errors in age and metallicity are not independent. In theory this
means that error bars in age and metallicity are not orthogonal and
the effect of either needs to be considered whenever plotting the
other. This could also lead to a false (negative) correlation in the
age-metallicity relation. In our sample, however, this must be a small
effect as log Age vs \feh \ is positively correlated (metallicity is
generally lower for younger galaxies). The typical error locus is
$\approx 1/20$ of the dynamic range. This indicates that the
co-dependence of these errors does not significantly distort the
derived physical properties of our sample. We find that the errors in
age and metallicity are independent of those of \afe.

\subsection{Emission line measurements}

Five galaxies have emission in at least two of the H$\beta$, [O {\sc
iii}] [N{\sc~ii}] or H$\alpha$ lines (see Table~\ref{tab:spe}). In two
cases, NGC~1581 and NGC~3156, this emission is spatially
resolved. According to the line-strength diagnostics of Kewley \etal
(2001)\nocite{kewley01}, NGC~2902 and NGC~1553 have unambiguous AGN
emission. We are unable to distinguish between thermal and non-thermal
emission for the other galaxies, though in those with
spatially-resolved emission some contribution must come from
star-formation. NGC~1581 also has strong (thermal) emission at $\sim 1
\times r_e$ from the nucleus suggestive of a ``ring'' of on-going
star formation.

\section{Discussion}
\label{sec:dis}

\subsection{Galaxy color versus age and metallicity}

\begin{figure}

  \includegraphics[width=80mm]{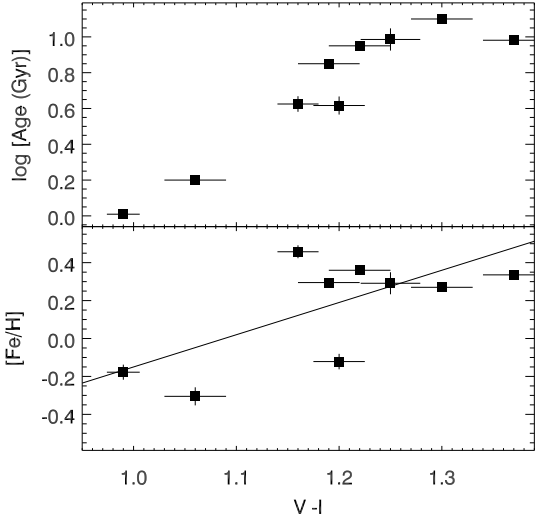}

  \caption{Luminosity-weighted mean age and metallicity vs total $V-I$
  color for those galaxies with quoted colors in Poulain \& Nieto
  (1994) and Prugniel \& Heraudeau (1998). The solid line in the
  bottom panel represents the transformation of the age-color and
  age-metallicity fits to the data to the metallicity-color
  plane.\label{fig:col}}\nocite{poulain94,prugniel98}

\end{figure}

Arag{\'o}n-Salamanca \etal (2006)\nocite{aragon-salamanca06} found the
relationship between color and $S_N$ for S0s is consistent with a
fading stellar population. However, because of the age-metallicity
degeneracy, they were unable to confirm that this was truly an age
effect. Now that we have separated age and \feh, we can tackle this
question directly. Figure~\ref{fig:col} shows $V - I$ color versus age
and metallicity. This plot clearly indicates that age is strongly
correlated with $V - I$ while metallicity is less so. A Spearman
rank-correlation test indicates that the $V-I$ log Age correlation is
significant at greater than the $99\%$ level, while that of
metallicity is $<94\%$. The variation of [Fe/H] with color can be
explained by the age-color relationship and the age-metallicity
degeneracy. On the bottom panel in Figure~\ref{fig:col} we plot the
linear fits to the our age-color and age-metallicity data transformed
to the metallicity-color plane. The residuals from this fit are not
correlated with color. The galaxies plotted here are a subset of those
in Arag{\'o}n-Salamanca \etal (2006)\nocite{aragon-salamanca06}. We
thus confirm that their result was correctly interpreted as an age
effect rather than being due to metallicity.

\subsection{The formation of S0 Galaxies}

\begin{figure}

  \includegraphics[width=80mm]{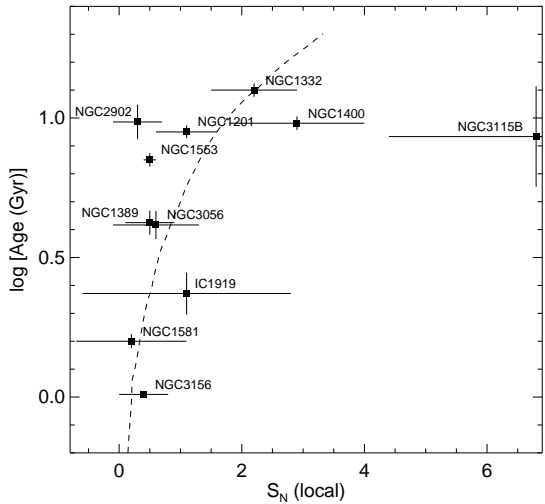}

  \caption{Log age in Gyr vs globular cluster specific frequency
  ($S_N$) for our sample. The line shows the evolution expected for a
  fading galaxy which starts out with $S_N = 0.4$ according to the
  stellar population models of Bruzual \& Charlot
  (2003).\label{fig:age}}\nocite{bruzual03}

\end{figure}

Figure~\ref{fig:age} shows the derived ages versus globular cluster
specific frequency. Also shown is the relationship expected from
population synthesis models of a fading stellar population (Bruzual \&
Charlot 2003)\nocite{bruzual03}. This relationship is normalised to a
spiral with $S_N = 0.4$. We note that younger S0s are found with $S_N$
close to the spiral values whereas older S0s are nearer the average
elliptical $S_N$.

Our results are consistent with those presented in
Arag{\'o}n-Salamanca \etal (2006)\nocite{aragon-salamanca06} and
support the theory that S0 galaxies evolve from fading spirals. It is
important to understand that the distribution of our points on the
Age-$S_N$ diagram is not an evolutionary sequence with one galaxies
moving from one point to another. Rather S0s will evolve differently
according to their mass and the total number of globular
clusters. However, the galaxies do occupy the locus of points expected
from the fading population model and occupy that space between the
average spiral $S_N$ and average elliptical $S_N$.

The fact that there are no young galaxies with large values of $S_N$
indicates that large numbers of globular clusters are not created when
S0s form. More precisely, any increase in the number of globular
clusters must be accompanied by an increase of the same factor in the
luminosity of the galaxy to stop $S_N$ increasing.

\subsection{Correlations between $M_V$, age, \afe, and $\sigma$}

A number of relationships are expected, and indeed found, for our
sample of galaxies. Before drawing any conclusions we need to
establish firmly that these do not affect our Age $- S_N$ plot in a
way that might mimic the effect of a fading stellar population. The
strongest correlations seen in our data are (1) the Faber-Jackson
relation ($M_V - \sigma$), (2) $M_V - $ \afe, and (3) $M_V - $log
Age. The first two relationships are well established (\eg Faber \&
Jackson 1976\nocite{faber76}; J{\o}rgensen 1999\nocite{jorgensen99};
Trager \etal 2000\nocite{trager00}), and the latter two are a
manifestation of the ``downsizing'' phenomenon. Qualitatively this
says that massive galaxies are older and formed more quickly (see \eg
Cowie \etal 1996\nocite{cowie96}; Gavazzi \etal
1996\nocite{gavazzi96}; Boselli \etal
2001\nocite{boselli01}). Relationship (3) will have the most direct
effect on our Age $- S_N$ plot as $M_V$ enters the definition of $S_N$
directly. For the galaxies in our sample $\log \mathrm{Age} = -0.326
M_V - 5.73$.

The $M_V - $log Age relation cannot mimic the fading of a stellar
population on the Age $- S_N$ diagram. In the former case, older
galaxies are brighter and so will tend to have lower $S_N$. In a
fading population the opposite is true. The reality will be slightly
more complex as brighter galaxies tend to have more clusters. However,
it is clear that the $M_V - $log Age relationship acts in a different
direction and so cannot be responsible for our Age $- S_N$
distribution.

\subsection{On-going star formation}

It has been suggested that an environment-induced transformation from
spiral to S0 may be accompanied by a starburst (\eg Milvang-Jensen
\etal 2003; Dressler \etal 2004; Bamford \etal
2005)\nocite{milvang-jensen03,dressler04,bamford05}. The two youngest
galaxies in our sample do indeed show signs of on-going
star-formation. The rest of the galaxies have luminosity-weighted mean
ages greater than 2 Gyr. This is probably longer than any
environment-induced burst would last and so we would not reasonably
expect to see forming stars in these galaxies. The pertinent question
is whether the star-formation in the youngest galaxies can be traced
to a burst $\sim 1$ Gyr in the past, or represents the last vestiges
of normal spiral activity. The duration of a starburst can be expected
to be of order a group-or-cluster crossing time ($\lesssim 10^8$ yr),
and we predict that it will use up the galactic gas reservoir more
quickly than normal star formation. It therefore seems more likely
that current emission lines represent the dying embers of regular
spiral activity. The fact of on-going star-formation also argues
against a rapid method of creation for lenticulars. It is difficult to
envision star formation lasting $\simeq 1$ Gyr for any scenario
involving close galaxy interactions or mergers. However, a complete
rejection of a violent formation scenario using this argument is
premature given that we are speculating on the nature of line emission
in just two galaxies.

\subsection{Comparison with other studies}

A number of methods of transforming spirals to lenticulars have been
proposed (see Introduction). As mentioned before, however,
observational studies, rather than plausible simulations are less
numerous. Studies of rich clusters at intermediate redshift show that
S0s must have undergone starbursts in the not too distant past (\eg
Treu \etal 2003; Dressler \etal 2004; McIntosh \etal 2004; Barr \etal
2006)\nocite{treu03,dressler04,mcintosh04,barr06}. Other studies
favouring the ``violent'' formation scenario show tidally-induced star
formation (\eg Boselli et al 2005\nocite{boselli05}) or
ram-pressure-induced star formation (\eg Sun \etal
2006\nocite{sun06}).

A gentler transformation from spiral to S0 is implied by the studies
of Bedregal \etal 2006 and Boselli \etal
2006\nocite{bedregal06,boselli06}. Our observations are consistent
with fading of the stellar populations. This, combined with the
on-going star formation in our youngest galaxies leads the present
study toward the gentler formation mechanism.  It must be borne in
mind, however, that our galaxies generally reside in poorer
environments than those supporting the more rapid transformation
mechanisms.  This may signal an important distinction with starbursts
more common in richer environments and a study of the differences
between cluster, group and field S0s would address this issue. For the
moment, our sample does not allow us to draw conclusions in this area.

\subsection{NGC~3115B}

If our understanding of the evolution in the Age, $S_N$ plane is
correct then NGC~3115B shouldn't end up where it is. Kundu \& Whitmore
(2001b)\nocite{kundu01b} note that this is a ``cluster rich'' dwarf
galaxy and perhaps some of its larger neighbour's systems have been
mistakenly assigned to it. This would suggest that $S_N$ should be
lower than quoted. It is an outlier by over 3 magnitudes on the M$_V$
-- $S_N$ relation.

\section{Summary}
\label{sec:con}

We test the theory that S0 galaxies formed from spirals whose star
formation is shut off and who evolve passively from that moment. Using
EMMI on the NTT we have taken long slit spectroscopy of eleven
lenticular (S0) galaxies at $z < 0.006$. We derive absorption line
indices from the central $r_e/8$ of each galaxy. These measurements
are then used, together with stellar population models to estimate
relative ages, metallicities and $\alpha$-element abundance ratios.

The derived physical properties are compared with globular cluster
specific frequency ($S_N$). We find that $S_N$ is correlated with
age. The values of $S_N$ span the range of average values for spirals
to the average value of ellipticals. Our galaxies occupy the locus of
points on the Age$-S_N$ diagram expected for a stellar population
fading from the average spiral $S_N$. We confirm that previous results
showing a relationship between $S_N$ and color are driven by the $S_N
-$ Age effect.

Our results are consistent with the hypothesis that S0 galaxies are
formed from spirals. An individual galaxy's position on the Age $-
S_N$ plot is a function of the time since the cessation of star
formation. The two youngest galaxies in our sample have extended
emission lines indicating on-going star-formation. This points to a
timescale over which star formation is shut down which is of order 1
Gyr. We speculate that the truncation of star formation may therefore
be a gentle process involving few, if any, major bursts.

\begin{acknowledgements}

We thank the referee, Alessandro Boselli, for insightful comments
which improved this paper. MRM is supported by a PPARC Senior
Fellowship. Based on observations made with the NTT ESO telescope at
La Silla observatory under programme ID 076.B-0182(A)

\end{acknowledgements}

\end{document}

%% file: tab1.asc
NGC1201  &   3600 &        $-20.8$    &       1.1   &        0.5  &         28.00  &   0.005722  \\
NGC1332  &   2700 &        $-21.2$    &       2.2   &        2.7  &         28.00  &   0.005219  \\
NGC1389  &   3600 &        $-19.5$    &       0.5   &        0.4  &       15.04    &   0.003342    \\
NGC1400  &   4500 &        $-20.6$    &       2.9   &        1.1  &       29.32    &   0.001964   \\
NGC1553  &   2700 &        $-21.5$    &       0.5   &        0.1  &       65.63    &   0.004227    \\
NGC1581  &   7200 &        $-18.2$    &       0.2   &        0.9  &       10.64    &   0.005390    \\
IC1919   &  19800 &        $-18.2$    &       1.1   &        1.7  &         60.00  &   0.004511  \\
NGC2902  &   2700 &        $-20.2$    &       0.3   &        0.4  &       15.39    &   0.006611    \\
NGC3056  &   2700 &        $-18.9$    &       0.6   &        0.7  &       50.95    &   0.003229    \\
NGC3115B &  14400 &        $-17.9$    &       6.8   &        2.4  &       16.87    &   0.002316    \\
NGC3156  &   3600 &        $-18.9$    &       0.4   &        0.4  &       14.36    &   0.004396    \\

%% file: tab2.asc
NGC1201     &         111   &     3  &   $-7.26$  &      0.04  &   $ -2.19$    &    0.03  &    4.70   &    0.04  &    3.08 &   0.03             &  0.950 &   0.022  &   $ 0.360$   & 0.022 &    0.185  &      0.012 \\ 
NGC1332     &         240   &     4 &    $-7.02$  &      0.04  &   $ -1.76$    &    0.03  &    5.04   &    0.04  &    2.75 &   0.03             &  1.103 &   0.024  &   $ 0.270$   & 0.023 &    0.325  &      0.012 \\ 
NGC1389     &          82   &     2  &   $-6.08$  &      0.05  &   $ -1.75$    &    0.04  &    4.25   &    0.05  &    3.12 &   0.04             &  0.625 &   0.044  &   $ 0.467$   & 0.034 &    0.125  &      0.014 \\ 
NGC1400     &         215  &      7  &   $-6.62$  &      0.04  &   $ -1.36$    &    0.03  &    5.05   &    0.04  &    2.74 &   0.03             &  0.981 &   0.024  &   $ 0.345$   & 0.023 &    0.341  &      0.011  \\ 
NGC1553$^{\star}$     &         137   &     1 &    $-6.78$  &      0.03  &   $ -1.33$    &    0.02  &    4.33   &    0.03  &    2.94  &   0.03  &  0.850 &   0.025  &   $ 0.305$   & 0.024 &    0.175  &      0.011 \\ 
NGC1581$^{\star}$      &          74   &     5  &   $ 1.11$  &      0.07  &   $  4.70$    &    0.05  &    1.83   &    0.11  &    1.73 &   0.09  &  0.204 &   0.025  &   $-0.315$   & 0.005 &    0.075  &      0.056 \\ 
IC1919      &          76   &     1  &   $-2.32$  &      0.18  &   $  1.30$    &    0.12  &    2.42   &    0.30  &    2.57 &   0.23             &  0.371 &   0.075  &   $ 0.072$   & 0.175 &    0.016  &      0.027 \\ 
NGC2902$^{\star}$      &         108  &      1  &   $-6.66$  &      0.09  &   $ -2.12$    &    0.07  &    4.68   &    0.11  &    2.91 &   0.08  &  0.985 &   0.062  &   $ 0.291$   & 0.058 &    0.225  &      0.027 \\ 
NGC3056$^{\star}$      &          74   &     5  &   $-3.90$  &      0.06  &   $  0.80$    &    0.04  &    2.69   &    0.07  &    2.36 &   0.06  &  0.617 &   0.051  &   $-0.121$   & 0.041 &    0.025  &      0.025 \\ 
NGC3115B    &          66   &     1  &   $-3.81$  &      0.31  &   $  0.29$    &    0.22  &    2.81   &    0.43  &    2.03 &   0.33             &  0.933 &   0.180  &   $-0.321$   & 0.166 &    0.217  &      0.155 \\ 
NGC3156$^{\star}$      &          72   &     2  &   $ 3.95$  &      0.05  &   $  6.37$    &    0.03  &    1.72   &    0.07  &    1.76 &   0.06  &  0.010 &   0.009  &   $-0.187$   & 0.044 &    0.036  &      0.025 \\